\documentstyle[11pt,epsfig,axodraw,amsmath]{article}
\begin{document}
\begin{titlepage}
%\title{}
\begin{center}
{\Large\bf Local  $D3/D7$ $\mu$-Split SUSY, 125 $GeV$ Higgs and Large Volume Ricci-Flat Swiss-Cheese Metrics: A Brief Review}
\vskip 0.1in  Aalok Misra\footnote{e-mail: aalokfph@iitr.ernet.in
%; tel. no:+91-1332-285404;fax no:+91-1332-273560
}\\
Department of Physics, Indian Institute of Technology,
Roorkee - 247 667, Uttaranchal, India\\
%(b) Physics Department, Theory Unit, CERN, CH1211, Geneva 23, Switzerland } \vskip 0.5 true in
%\date{\today}
\end{center}
\thispagestyle{empty}

%\maketitle

%\pub{Received (Day Month Year)}{Revised (Day Month Year)}

\begin{abstract}
In this article, we  review briefly recent progress made in realizing local(ized around a mobile space-time filling $D3$-brane in) $D3/D7$ $\mu$-Split Supersymmetry in (the large volume limit of Type IIB) String Theory (compactified on Swiss-Cheese Calabi-Yau orientifolds) as well as obtaining a 125 $GeV$ (light) Higgs in the same set up. We also discuss obtaining the geometric K\"{a}hler potential (and hence the Ricci-Flat metric) for the Swiss-Cheese Calabi-Yau in the large volume limit using the Donaldson's algorithm and intuition from GLSM-based calculations - we present new results for Swiss-Cheese Calabi-Yau (used in the set up) metrics at points finitely away from the ``big" divisor.

%\keywords{$\mu$-Splift SUSY; Swiss-Cheese Calabi-Yau's; Gluino decays; Donaldson's algorithm; %Ricci-flat metrics.}
\end{abstract}
\end{titlepage}
%\ccode{PACS Nos.: 11.25.-w; 11.25.Sq; 11.15.Tk.}

\section{Introduction and Review of Setup}	

Despite the success of Standard Model in High Energy Physics, failure of naturalness and fine tuning requirements in the Higgs Sector remain basic motivations for constructing theories beyond the Standard Model. The supersymmetric extension(s) of the Standard Model  can solve the fine tuning problem in the Higgs/scalar sector, however for this one  requires supersymmetric particles at TeV scale. Though it is possible to achieve gauge coupling unification and obtain a  dark matter candidate, yet the existence of naturally large supersymmetric contribution to flavor changing neutral current, experimental value of electron dipole moment (EDM) for natural CP violating phase and dimension-five proton decays are serious issues that can not be solved elegantly in supersymmetric Standard Model. Also, lack of existence of light Higgs boson is one of the major tensions in MSSM. An alternative approach to SUSY was adopted by Arkani-Hamed and Dimopoulos\cite{HamidSplitSUSY} in which they argued given that fine tuning anyway seems to be required to obtain a small and positive cosmological constant (which is one of the most serious issues), one is hence also allowed to assume fine tuning in other sectors of the theory (Higgs Sector) which is a less serious issue in the string theory landscape.  Their model based on high scale $(m_s\sim {10}^{10}$ GeV) SUSY breaking is named as split SUSY Model. In this scenario all scalar particles acquire heavy masses except one Higgs doublet which is finely tuned to be light while fermions (possibly also gaugino and Higgsino) are light. This  interesting class of model has attracted considerable attention though it abandons the primary reason for introducing supersymmetry. This scenario removes all unrealistic features of MSSM while preserves all good features ( possibly gauge coupling unification and dark matter candidate).  In \cite{sudhir kumar gupta} it is shown that the lightest neutralino can still be taken as a good dark matter candidate in split SUSY. Also gauge coupling unification remains inherent in split SUSY see \cite{haba and okada}. One of the other striking feature of this model based on heavy squark masses is the issue of gluino decay discussed in \cite{gluino decay pattern}. Kinematically favored three body gluino decays $\tilde{g}\rightarrow{\chi_i}^0{\bar{q}_J}q_J$  or $\tilde{g}\rightarrow{\chi_i}^\pm{\bar{q}_I}q_J$ (where ${\chi_i}^0$ ,${\chi_i}^\pm$ correspond to neutralinos and charginos, $q_{I,J}, {\bar{q}_{I,J}}$ correspond to quarks and antiquarks) occurring via virtual squarks get considerably suppressed due to heavy squark masses and hence gluino remains long lived. Therefore measuring life time of gluino can be adopted as indirect way to measure heavy squark mass i.e limit of SUSY breaking scale in split SUSY scenario.

 Despite explaining many unresolved issues of phenomenology in the context of split SUSY, the notorius ${\mu}$ problem still
 remains unsolved according to which the stable vaccum that spontaneously breaks electroweak
 symmetry requires ${\mu}$ to be of the order of supersymmetry breaking scale.  However in case of split SUSY scenario one is assuming ${\mu}$ to be light while
 supersymmetry breaking scale to be very high. The other alternative  to solve the $\mu$ problem has been discussed by authors in \cite{mu split susy} in which they introduce a further split in the split SUSY scenario by raising the
 ${\mu}$ parameter to a large value which could be about the same as the sfermion mass or the SUSY breaking scale; this scenario is dubbed as ${\mu}$-split SUSY scenario. In addition to solving the $\mu$ problem, all the nice features of split
 supersymmeric model like gauge coupling unification, dark matter candidate remain protected in
 this scenario.

With the promising approach of string theory to phenomenology as well as cosmology, it is quite interesting to realize the split SUSY scenario within a string theoretic framework. The signatures of the same in the context of type I and type IIA string theory were obtained respectively in \cite{type 1 string} and \cite{Kokorelis}. Recently, in the context of type IIB (``big divisor") LVS $D3/D7$ Swiss cheese phenomenology, the authors of \cite{ferm_masses_MS} explicitly showed the possibility of generating light fermion masses as well as heavy squark/sleptons masses    including a space-time filling mobile $D3$ brane and stack(s) of (fluxed) $D7$- branes wrapping the ``Big" divisor. Matter fields (quarks, leptons and their superpartners) are identified with the (fermionic superpartners of) Wilson line moduli whereas Higgses are identified with space-time filling mobile $D3$-brane position moduli.

In the remainder of this section, we next briefly describe our setup: type IIB compactification on the orientifold of a ``Swiss-Cheese Calabi-Yau" in the large volume limit including perturbative $\alpha^\prime$ and world sheet instanton corrections
%as well as one-loop corrections
to the K\"{a}hler potential, and the instanton-generated superpotential written out respecting the (subgroup, under orientifolding, of) $SL(2,{\bf Z})$ symmetry of the underlying parent type IIB theory, localized around a mobile space-time filling $D3-$brane ``restricted" to stacks of $D7$-branes wrapping the ``big" divisor along with magnetic fluxes. This is followed by a summary of evaluation of soft supersymmetry breaking parameters, showing that one obtains large open string moduli
masses; based on the Yukawas calculated and summarized in Table 1, one conjectures that the $D3$-brane position moduli could be identified with the Higgses and the Wilson line moduli with
the first two generations' squarks/sleptons.

%the possibility of getting light fermions and heavy scalar superpartners and generating (less than) $eV$ mass scales relevant to Majorana neutrino mass scales.

In  \cite{dSetal,largefNL_r_axionicswisscheese}, we  addressed some cosmological issues like $dS$ realization, embedding inflationary scenarios and realizing non-trivial non-Gaussianities in the context of type IIB Swiss-Cheese Calabi Yau orientifold in LVS. This has been done with the inclusion of (non-)perturbative $\alpha^{\prime}$-corrections to the K\"{a}hler potential and non-perturbative instanton contribution to the superpotential. The Swiss-Cheese Calabi Yau we are using, is a projective variety in ${\bf WCP}^4[1,1,1,6,9]$ given as
\begin{equation}
\label{eq:hyper}
x_1^{18} + x_2^{18} + x_3^{18} + x_4^3 + x_5^2 - 18\psi \prod_{i=1}^5x_i - 3\phi x_1^6x_2^6x_3^6 = 0,
\end{equation}
 which has two (big and small) divisors $\Sigma_B(x_5=0)$ and $\Sigma_S(x_4=0)$\cite{denef_LesHouches}. From Sen's orientifold-limit-of-F-theory point of view  corresponding to type IIB compactified on a Calabi-Yau three fold $Z$-orientifold with $O3/O7$ planes, one requires a Calabi-Yau four-fold $X_4$ elliptically fibered (with projection $\pi$) over a 3-fold $B_3(\equiv CY_3-$orientifold)  where $B_3$ is taken be  an $n$-twisted ${\bf CP}^1$-fibration over ${\bf CP}^2$ such that pull-back of the divisors in $CY_3$ automatically satisfy Witten's unit-arithmetic genus condition.  For $n=6$ \cite{DDF}, the $CY_4$ will be the resolution of a Weierstrass model with $D_4$ singularity along the first section and an $E_{6/7/8}$ singularity along the second section. The Calabi-Yau three-fold $Z$ then turns out to be a unique Swiss-Cheese Calabi Yau  in ${\bf WCP}^4[1,1,1,6,9]$ given by (\ref{eq:hyper}). We would be assuming an $E_8$-singularity as this corresponds to
 $h^{1,1}_-(CY_3)=h^{2,1}(CY_4)\neq0$\cite{denef_LesHouches} which is what we will be needing and using. The required Calabi-Yau has $h^{1,1}=2, h^{2,1}=272$. The same has a large discrete symmetry group given by $\Gamma={\bf Z}_6\times{\bf Z}_{18}$ (as mentioned in \cite{D3_D7_Misra_Shukla}) relevant to construction of the mirror a la Greene-Plesser prescription. However, as is common in such calculations, one assumes that one is working with a subset of periods of $\Gamma$-invariant cycles - the six periods corresponding to the two complex structure deformations in (\ref{eq:hyper}) will coincide with the six periods of the mirror - the complex structure moduli absent in (\ref{eq:hyper}) will appear only at a higher order in the superpotential because of $\Gamma$-invariance and can be consistently set to zero.

As shown in \cite{D3_D7_Misra_Shukla}, in order to support MSSM (-like) models and for resolving the tension between LVS cosmology and LVS phenemenology within a string theoretic setup, a mobile space-time filling $D3-$brane and stacks of $D7$-branes wrapping the ``big" divisor $\Sigma_B$ along with magnetic fluxes, are included. The appropriate ${\cal N}=1$ coordinates in the presence of a single $D3$-brane and a single $D7$-brane wrapping the big divisor $\Sigma^B$ along with $D7$-brane fluxes were obtained in \cite{jockersetal}; the same along with the details of the holomorphic isometric involution involved in orientifolding, as well as  expansion of the complete K\"{a}hler potential (including  the geometric K\"{a}hler potential) and the (non-perturbative) superpotential as a power series in fluctuations about Higgses' vevs and the corresponding  extremum values of the Wilson line moduli, have been summarized in \cite{ferm_masses_MS}.

%Now, in the context of intersecting brane world scenarios \cite{int_brane_SM,Ibanez et al}, bifundamental leptons and quarks are obtained respectively from open strings stretched between $U(2)$ and $U(1)$ stacks, and $U(3)$ and $U(2)$ stacks of $D7$-branes; the adjoint gauge fields correspond to open strings starting and ending on the same $D7$-brane.
In Large Volume Scenarios, one considers four stacks of different numbers of multiple $D7$-branes  wrapping $\Sigma_B$ but with different  choices of magnetic $U(1)$ fluxes turned on, on the two-cycles which are non-trivial in the Homology of $\Sigma_B$ and not the ambient Swiss Cheese Calabi-Yau. By turning on different $U(1)$ fluxes on, e.g., the $3_{QCD}+2_{EW}$ $D7$-brane stacks in the LVS setup, $U(3_{QCD}+2_{EW})$ is broken down to $U(3_{QCD})\times U(2_{EW})$ and the four-dimensional Wilson line moduli $a_{I(=1,...,h^{0,1}_-(\Sigma_B))}$ and their fermionic superpartners $\chi^I$ that are valued, e.g., in the $adj(U(3_{QCD}+2_{EW}))$ to begin with, decompose into the bifundamentals $(3_{QCD},{\bar 2}_{EW})$ and its complex conjugate, corresponding to the bifundamental left-handed quarks of the Standard Model (See \cite{bifund_ferm}).  The inverse gauge coupling constant squared for the $j$-th gauge group ($j:SU(3), SU(2),U(1)$), up to open string one-loop level, using \cite{Maldaetal_Wnp_pref,Wilson 1loop,Jockers_thesis}, will be given by
\begin{eqnarray}
\label{eq:1overgsquared}
 \frac{1}{g_{j{=SU(3)\ {\rm or}\ SU(2)}}^2}&  = & Re(T_{S/B}) + ln\left(\left.P\left(\Sigma_S\right)\right|_{D3|_{\Sigma_B}}\right) + ln\left(\left.{\bar P}\left(\Sigma_S\right)\right|_{D3|_{\Sigma_B}}\right)\nonumber\\
 & & + {\cal O}\left({\rm U(1)-Flux}_j^2\right),
\end{eqnarray}
 where
${\rm U(1)-Flux}_j$ are abelian magnetic fluxes for the $j-$th stack. Also, $\left.P\left(\Sigma_s\right)\right|_{D3|_{\Sigma_B}}$ implies the defining hypersurface for the small divisor $\Sigma_S$ written out in terms of the position moduli of the mobile $D3$-brane, restricted to the big divisor $\Sigma_B$. Further, the main idea then behind realizing $O(1)$ gauge coupling is the competing contribution as compared to the volume of the big divisor $\Sigma_B$ to the gauge kinetic function (and hence to the gauge coupling) coming from the $D7$-brane Wilson line moduli contribution $c_{I{\bar J}}a^I{\bar a}^{\bar J}$ where the intersection matrix $c_{I{\bar J}}=\int_{\Sigma_B}i^*\omega_B\wedge A_I\wedge {\bar A}_{\bar J}$ (the immersion map $i$ being defined as
$i:\Sigma^B\hookrightarrow CY_3$) and $\omega_B\in H^{1,1}_+$ - the Poincare-dual of $\Sigma_{B}$, i.e.,  $\omega_B=\delta (P_{\Sigma_B})dP_{\Sigma_B}\wedge \delta({\bar P}_{\Sigma_B})d{\bar P}_{\Sigma_B}$ (See \cite{denef_LesHouches}) - noting that $i_B^*dz_3\sim\frac{\phi z_1^5z_2^5(z_2dz_1+z_1dz_2)-(z_1^{17}dz_1+z_2^{17}dz_2)}{(\phi z_1^6z_2^6-z_1^{18}-z_2^{18}-1)^{\frac{2}{3}}}(i_B:\Sigma_B(1+z_1^{18}+z_2^{18}+z_3^3=\phi z_1^6z_2^6)\hookrightarrow CY_3)$, near $z_{1,2}\sim\frac{{\cal V}^{\frac{1}{36}}}{\sqrt{2}}$ (implying $dz_3\sim{\cal V}^{\frac{5}{36}}(dz_1+dz_2)$), is given by: $\omega_B\sim\frac{{\cal V}^{\frac{17}{18}}}{2^{17}}(dz_1+dz_2)\wedge(d{\bar z}_1+d{\bar z}_2)\biggr|_{{\cal V}\sim10^6}\sim (dz_1+dz_2)\wedge(d{\bar z}_1+d{\bar z}_2)$. After constructing the following local (i.e. localized around the location of the mobile $D3$-brane in the Calabi-Yau) appropriate involutively-odd harmonic distribution one-forms on the big divisor that lie in $coker\left(H^{(0,1)}_{{\bar\partial},-}(CY_3)\stackrel{i^*}{\rightarrow}
H^{(0,1)}_{{\bar\partial},-}(\Sigma_B)\right)$:
\begin{equation}
\label{eq:har one forms}
A_I\sim \delta\left(|z_3|-{\cal V}^{\frac{1}{6}}\right)\delta\left(|z_1|-{\cal V}^{\frac{1}{36}}\right)\delta\left(|z_2|-{\cal V}^{\frac{1}{36}}\right)\left[\omega_I(z_1,z_2)dz_1+\tilde{\omega}_I(z_1,z_2)dz_2\right],
\end{equation}
where $\omega(-z_1,z_2)=\omega(z_1,z_2), \tilde{\omega}(-z_1,z_2)=-\tilde{\omega}(z_1,z_2)$ and $\partial_1\tilde{\omega}=-\partial_2\omega$ (for the large volume holomorphic isometric involution \cite{ferm_masses_MS}$\sigma: z_1\rightarrow -z_1, z_{2,3}\rightarrow z_{2,3}$); one obtains (See \cite{D3_D7_Misra_Shukla,ferm_masses_MS}):
\begin{eqnarray}
\label{eq:A_123}
& & \hskip -0.5in A_1(z_1,z_2,z_3\sim{\cal V}^{\frac{1}{6}})\sim - z_1^{18}z_2^{19}dz_1 + z_1^{19}z_2^{18}dz_2,\nonumber\\
& & \hskip -0.5in A_2(z_1,z_2,z_3\sim{\cal V}^{\frac{1}{6}})\sim - \left(\frac{z_2^{19}}{19}+z_1^{18}z_2\right)dz_1 + \left(\frac{z_1^{19}}{19}+z_2^{18}z_1\right)dz_2.
\end{eqnarray}
 This also involves stabilization of the Wilson line moduli at around $ {\cal V}^{-\frac{2}{9}}$ and the
  $D3$-brane position moduli, the Higgses in our setup, at around ${\cal V}^{\frac{1}{36}}$; extremization of the ${\cal N}=1$ potential, as shown in \cite{D3_D7_Misra_Shukla} and mentioned earlier on, shows that this is indeed true.  This way the gauge couplings corresponding to the gauge theories living on stacks of $D7$ branes wrapping the ``big" divisor $\Sigma_B$ (with different $U(1)$ fluxes on the two-cycles inherited from $\Sigma_B$) will be given by:
$g_{YM}^{-2}\sim {\cal V}^{\frac{1}{18}}$, $T_B$ being the appropriate ${\cal N}=1$ K\"{a}hler coordinate  and the relevant text below the same) and $\mu_3$ related to the $D3$-brane tension,
implying a finite (${\cal O}(1)$) $g_{YM}$ for ${\cal V}\sim10^6$. In the dilute flux approximation, the ``ln" terms in the right hand side of (\ref{eq:1overgsquared})are of ${\cal O}\left( ln{\cal V}\right)$, which for ${\cal V}\sim10^6$ is taken to be of the same order as $\sigma^B$(Big divisor's volume complexified by four-form axion)$+{\bar\sigma}^{\bar B} - C_{I{\bar J}}a^I{\bar a}^{\bar J}\sim{\cal V}^{\frac{1}{18}}$ appearing in $Re(T_B)$. %For ${1}/{g_{U(1)}^2}$ there is a model-dependent numerical prefactor multiplying the right hand side of the $1/g_j^2$-relation.
In the dilute flux approximation, $\alpha_i(M_s)/\alpha_i(M_{EW}), i=SU(3),SU(2),U(1)_Y$, are hence unified.

As discussed in \cite{Sparticles_Misra_Shukla}, for the type IIB Swiss-Cheese orientifold considered in our work, guided, e.g., by the vanishingly small Yukawa couplings $\hat{Y}_{\tilde{\cal A}_1^2{\cal Z}_i}\equiv\frac{e^{\frac{K}{2}}Y_{\tilde{\cal A}_1^2{\cal Z}_i}}{\sqrt{\left(K_{\tilde{\cal A}_1\bar{\tilde{{\cal A}}}_1}\right)^2K_{{\cal Z}_i{\bar{\cal Z}_i}}}}$\footnote{The Yukawa couplings, $K_{{\cal C}_i\bar{\cal C}_{\bar j}}$, etc. are defined by taking derivatives of the superpotential and K\"{a}hler potential with respect to fluctuations ${\cal C}_{i,1}\equiv\delta {\cal Z}_i,\delta\tilde{\cal A}_1$ in the open string moduli (about non-zero vevs) written in the basis: $\delta{\cal Z}_i=\delta z_i + {\cal O}(1){\cal V}^{-\frac{8}{9}}\delta {\cal A}_1, \delta\tilde{{\cal A}}_1=\delta{\cal A}_1+{\cal V}^{-\frac{8}{9}}\left({\cal O}(1)\delta z_1 + {\cal O}(1)\delta z_2\right)$, which diagonalizes $K_{{\cal C}_i\bar{\cal C}_{\bar j}}$\cite{D3_D7_Misra_Shukla}.} obtained from an  $ED3$-instanton-generated superpotential (See Table 1 for the single Wilson line modulus case), the spacetime filling mobile D3-brane position moduli $z_i$'s and the Wilson line moduli $a_{I}$'s could be respectively identified with Higgses and the either of the first two generations of sparticles (squarks/sleptons) of some (MS)SM-like model.
With a (partial) cancelation between the volume of the ``big" divisor and the Wilson line
contribution (required for realizing $\sim O(1) g_{YM}$ in our setup), in \cite{D3_D7_Misra_Shukla}, we calculated in the large volume limit, several soft supersymmetry breaking parameters. The same relevant to this review are summarized in table 1.
 \begin{table}[htbp]
\centering
\begin{tabular}{|l|l|}
\hline
Gravitino mass &  $ m_{\frac{3}{2}}\sim{\cal V}^{-\frac{n^s}{2} - 1}$ \\
Gaugino mass & $ M_{\tilde g}\sim m_{\frac{3}{2}}$\\ \hline
$D3$-brane position moduli  & $ m_{{\cal Z}_i}\sim {\cal V}^{\frac{19}{36}}m_{\frac{3}{2}}$ \\
(Higgs) mass & \\
Wilson line moduli mass & $ m_{\tilde{\cal A}_1}\sim {\cal V}^{\frac{73}{72}}m_{\frac{3}{2}}$\\ \hline
%& $A_{{\cal Z}_i{\cal Z}_j{\cal Z}_k}\sim n^s{\cal V}^{\frac{37}{36}}m_{\frac{3}{2}}$\\
A-terms & $A_{pqr}\sim n^s{\cal V}^{\frac{37}{36}}m_{\frac{3}{2}}$\\
& $\{p,q,r\} \in \{{{\tilde{\cal A}_1}},{{\cal Z}_i}\}$\\
%& $A_{{\tilde{\cal A}_1}{{\cal Z}_i}{{\cal Z}_j}}\sim n^s{\cal V}^{\frac{37}{36}}m_{\frac{3}{2}}$\\
\hline
Physical $\mu$-terms & $\hat{\mu}_{{\cal Z}_i{\cal Z}_j}\sim{\cal V}^{\frac{37}{36}}m_{\frac{3}{2}}$ \\
& $\hat{\mu}_{{\cal A}_1{\cal Z}_i}\sim{\cal V}^{-\frac{3}{4}}m_{\frac{3}{2}}$\\
& $\hat{\mu}_{{\cal A}_1{\cal A}_1}\sim{\cal V}^{-\frac{33}{36}}m_{\frac{3}{2}}$\\
\hline
%& $\hat{Y}_{{\cal Z}_i{\cal Z}_i{\cal Z}_i}\sim {\cal V}^{\frac{43}{24}}m_{\frac{3}{2}}$\\
%& (NMSSM)\\
%& $\hat{Y}_{{\cal Z}_i^2{\cal Z}_j}\sim {\cal V}^{\frac{37}{24}}m_{\frac{3}{2}}$\\
%& (NMSSM)\\
Physical Yukawa couplings&
%$\hat{Y}_{{\cal Z}_i^2\tilde{\cal A}_1}\sim {\cal V}^{\frac{1}{72}}m_{\frac{3}{2}}$\\
$\hat{Y}_{{\cal Z}_1{\cal Z}_2\tilde{\cal A}_1}\sim {\cal V}^{-\frac{17}{72}}m_{\frac{3}{2}}$\\
&$\hat{Y}_{\tilde{\cal A}_1^2{\cal Z}_i}\sim {\cal V}^{-\frac{127}{72}}m_{\frac{3}{2}}$\\
&$\hat{Y}_{\tilde{\cal A}_1\tilde{\cal A}_1\tilde{\cal A}_1}\sim {\cal V}^{-\frac{85}{24}}m_{\frac{3}{2}}$\\
\hline
%&$\left(\hat{\mu}B\right)_{{\cal Z}_i{\cal Z}_i}\sim{\cal V}^{\frac{223}{108}}m_{\frac{3}{2}}^2$\\
Physical $\hat{\mu}B$-terms & $\left(\hat{\mu}B\right)_{{\cal Z}_1{\cal Z}_2}\sim{\cal V}^{\frac{37}{18}}m_{\frac{3}{2}}^2$\\
% &$\left(\hat{\mu}B\right)_{\tilde{\cal A}_1\tilde{\cal A}_1}\sim{\cal V}^{\frac{5}{36}}m_{\frac{3}{2}}^2$\\
%&$\left(\hat{\mu}B\right)_{{\cal Z}_i\tilde{{\cal A}_1}}\sim{\cal V}^{-\frac{13}{18}}m_{\frac{3}{2}}^2$\\
\hline
\end{tabular}
\caption{{\small Results on Soft SUSY Parameters Summarized}}
\end{table}

\section{Obtaining Big Divisor $D3/D7$ $\mu$-Split SUSY as well as 125 GeV Higgs}

In split supersymmetry scenario, SUSY breaking scale is high. However, in order to get one light
Higgs doublet at EW scale in this scenario, one needs these soft terms to be of $TeV$ order. Since fine tuning is allowed one can assume ${{\hat\mu}B}\sim m_{{\cal A}_I}^2$ (where $m_{{\cal A}_I}$'s correspond to squark/slepton masses scale which is of the order of high supersymmetry breaking
scale as in case of split SUSY, and $\hat{\mu}_{{\cal Z}_1{\cal Z}_2}$ is the Higgsino mass
parameter). As Higgsino mass contribution ($\hat{\mu}_{{\cal Z}_1{\cal Z}_2}$ parameter) is small
in most of split SUSY models, one needs $B >>\hat{\mu}_{{\cal Z}_1{\cal Z}_2}$ in order to have
$\hat\mu{B}\sim{m^2_{{\cal A}_I}}$. In an alternate approach to split SUSY scenario called
``$\mu$-split SUSY scenario"  \cite{mu split susy}, according to which one can assume even
$\hat\mu\sim {m_{{\cal A}_I}}\sim B$ i.e large $\mu$ parameter to get ${{\hat\mu}B}\sim{m^2_{{\cal A}_I}}$, this choice appears more natural and also helps to alleviate the $``{\mu}$ problem"; see
also \cite{Split SUSY mu}. In the single-Wilson-line modulus Large Volume Scenarios set up discussed earlier, values of
$\hat\mu$ and $B$ terms pertaining to SUSY breaking parameters has been summarized in results
in\cite{D3_D7_Misra_Shukla}, which are of the order ${{\hat\mu}^2}\sim{\hat\mu{B}}\sim{m^2_{{\cal
A}_I}}$ (scalar masses) i.e ${\hat\mu}\sim{B}\sim{m_{{\cal A}_I}}$ as in case of $\mu$ split SUSY.

In this section, we demonstrate the possibility to realize $\mu$-Split SUSY  in the framework summarized in section {\bf 1}. We do so by first summarizing our results of \cite{D3_D7_Misra_Shukla,ferm_masses_MS,Dhuria+Misra_mu_Split SUSY} wherein we had shown that we obtain very heavy squarks/sleptons, heavy Higgsino mass parameter, light fermions and one light Higgs whose mass could be fine tuned to the desirable $125 GeV$.

\subsection{Generation of $\mu$-Split SUSY Mass Scales for Scalars and Fermions}

Fermion (Quark/Lepton) masses are generated by giving some VEVs to Higgses in $\int d^4 x \,e^{\hat{K}/2} Y_{ijk} {z}^{i} \psi^{j}\psi^{k}$. The (canonically normalized) fermionic mass matrix is generated by ${\hat {Y}_{ijk}<{z}_i }>$. For the single Wilson line modulus case, the  mass of the fermionic superpartner of $\tilde{\cal A}_1$, as shown in \cite{ferm_masses_MS} (which based on the near-vanishing value of the Yukawa coupling $\hat{Y}_{\tilde{\cal A}_1^2{\cal Z}_i}$ in Table 1, is conjectured to be a first/second generation quark/lepton), turns out to be given by:${\cal V}^{-\frac{199}{72}-\frac{n^s}{2}}$ in units of $M_p$, which  implies a range of fermion mass $m_{\rm ferm}\sim{\cal O}({\rm MeV-GeV})$ for Calabi Yau volume ${\cal V}\sim {\cal O}(6\times10^5-10^5)$. Interestingly, the mass-scale of $0.5$ MeV- the electronic mass scale- could be realized with ${\cal V}\sim 6.2\times 10^5, n^s=2$.  In MSSM/2HDM models, up to one loop, the leptonic (quark) masses do not change (appreciably) under an RG flow from the intermediate string scale down to the EW scale (See \cite{Das_Parida}).

The non-zero neutrino masses are generated through the Weinberg(-type) dimension-five operators written out schematically as: $\int d^4x\int d^2\theta e^{\hat{K}/2}\times\left({\cal Z}^2{\cal A}_1^2\in\frac{\partial^2{\cal Z}^4}{\partial {\cal Z}^2}{\cal A}_1^2\right)$,  and is given as:
$m_{\nu}={v^2 sin^2\beta \hat{{\cal O}}_{{\cal Z}_i{\cal Z}_j{\cal Z}_k{\cal Z}_l}}/{2M_p}
$ where $\hat{{\cal O}}_{{\cal Z}_i{\cal Z}_i{\cal Z}_i{\cal Z}_i}\equiv$coefficient of the physical/normalized quartic in  ${\cal Z}_i$ in the superpotential,  and is given as
$\hat{{\cal O}}_{{\cal Z}_i{\cal Z}_i{\cal Z}_i{\cal Z}_i}={\frac{{e^\frac{\hat{K}}{2}}{\cal O}_{{\cal Z}_i{\cal Z}_j{\cal Z}_k{\cal Z}_l}}{{\sqrt{\hat{K}_{{\cal Z}_i{\bar{\cal Z}}_{\bar i}}\hat{K}_{{\cal Z}_j{\bar{\cal Z}}_{\bar j}}\hat{K}_{{\cal Z}_k{\bar{\cal Z}}_{\bar k}}\hat{K}_{{\cal Z}_l{\bar{\cal Z}}_{\bar l}}}}}}$ \cite{conlon_neutrino},
$v sin\beta\equiv\langle H_u\rangle$ and $sin\beta$ is defined via
$tan\beta={\langle H_u\rangle}/{\langle H_d\rangle}$; in our setup (See \cite{ferm_masses_MS}):
$
 {\cal O}_{{\cal Z}_i{\cal Z}_j{\cal Z}_k{\cal Z}_l}\sim \frac{2^{n^s}}{24}10^2\left(\mu_3 n^s (2\pi \alpha^\prime)^2\right)^4{\cal V}^{\frac{n^s}{2}+\frac{1}{9}}
e^{-n^s{\rm vol}(\Sigma_s)+in^s\mu_3(2\pi\alpha^\prime)^2{\cal V}^{\frac{1}{18}}(\alpha+i\beta)}
$.
 Now, $z_i\sim\alpha_i{\cal V}^{\frac{1}{36}}, i=1,2; \beta\sim\alpha_1\alpha_2$ and
${\rm vol}(\Sigma_S)=\gamma_3 ln {\cal V}$ such that
$\gamma_3 ln{\cal V} + \mu_3l^2\beta {\cal V}^{\frac{1}{18}} = ln {\cal V}$, along with
$\hat{K}_{{\cal Z}_i{\bar{\cal Z}}_{\bar i}}\sim\frac{{\cal V}^{\frac{1}{72}}}
{\sqrt{\sum_\beta n^0_\beta}}$, and the assumption that the holomorphic isometric involution $\sigma$ as part of
the Swiss-Cheese orientifolding action $(-)^{F_L}\Omega\cdot \sigma$ is such that
$\sum_\beta n^0_\beta\sim\frac{{\cal V}}{{\cal O}(1)}$. By analying the RG running of coefficient $\kappa_{ij}$ of dimension-five operator $\kappa_{ij}L_iH.L_jH$ and $\langle H_u\rangle$, it was shown in \cite{ferm_masses_MS} that one can generate a neutrino mass of $\stackrel{<}{\sim}1eV$ in our setup.

%\subsection{Obtaining a 125 GeV Higgs and Heavy Higgsino}

We now summarize our calculations in \cite{Dhuria+Misra_mu_Split SUSY} wherein we had shown that the eigenvalues of the Higgs mass matrix at the EW scale obtained from the solutions to the one-loop RG flow equations assuming non-universality in the open string moduli masses, results in an eigenvalue corresponding to the mass-squared of one of the Higgs doublet to be negative and small and the other to be large and positive with a heavy Higgsino (in addition to heavy squarks/sleptons and light quarks/leptons already demonstrated in \cite{D3_D7_Misra_Shukla,Sparticles_Misra_Shukla,ferm_masses_MS}) implying the existence of $D3/D7$ $\mu$-Split LVS.

Due to lack of universality in moduli masses but universality in trilinear $A_{ijk}$ couplings, we need to use solution of RG flow equation for moduli masses as given in \cite{Nath+Arnowitt}.
From \cite{D3_D7_Misra_Shukla}, it was shown in \cite{Dhuria+Misra_mu_Split SUSY} that:
\begin{equation}
\label{eq:heavy_H_II}
\hskip-0.1in m_{{\cal Z}_1}^2(M_{EW})\sim m_{{\cal Z}_1}^2(M_s)+{(0.39)}m_{3/2}^2+\frac{1}{22}\times\frac{19\pi}{100}\times S_0,
\end{equation}
where $ S_0=Tr(Ym^2)=m_{{\cal Z}_2}^2-m_{{\cal Z}_1}^2+\sum_{i=1}^{n_g}(m_{\tilde q_L}^2-2
m_{\tilde u_R}^2 +m_{\tilde d_R}^2 - m_{\tilde l_L}^2 + m_{\tilde e_R}^2)$
in which all the masses are at the string scale and
 $n_g$ is the number of generations;
\begin{equation}
\label{eq:light_H_II}
m_{{\cal Z}_2}^2(M_{EW})\sim m_0^2\delta_2+{(0.32)}m_{3/2}^2+{(-0.03)}n^s\hat{\mu}_{{\cal Z}_1{\cal Z}_2}m_{3/2}+{(0.96)} m_0^2-{(0.01)}(n^s)^2\hat{\mu}_{{\cal Z}_1{\cal Z}_2}-\frac{19\pi}{2200}\times S_0,
\end{equation}
where we used $A_{{\cal Z}_i{\cal Z}_i{\cal Z}_i}\sim n^s\hat{\mu}_{{\cal Z}_1{\cal Z}_2}$ (See \cite{D3_D7_Misra_Shukla}).
The solution for RG flow equation for $\hat{\mu}^2$ to one loop order is given by \cite{Nath+Arnowitt}:
\begin{eqnarray}
\label{eq:muhat_I}
& & \hat{\mu}^2_{{\cal Z}_i{\cal Z}_i}=-\biggl[m_0^2 C_1+A_0^2 C_2 +m_{\frac{1}{2}}^2C_3+m_{\frac{1}{2}}
A_0C_4-\frac{1}{2}M_Z^2 +\frac{19\pi}{2200}\left(\frac{tan^2\beta+1}{tan^2\beta-1}\right)S_0\biggr],\nonumber\\
& &
\end{eqnarray}
where $C_{1,2,3,4}$ are as given in \cite{Nath+Arnowitt}. The overall minus sign on the right hand side of (\ref{eq:muhat_I}) indicates that our $\hat{\mu}_{{\cal Z}_1{\cal Z}_2}^2$ is negative of $\mu^2$ of \cite{Nath+Arnowitt}. In the large $tan\beta$ (but less than 50)-limit one sees that:
\begin{equation}
\label{eq:muhat_II}
\hskip -0.3in\hat{\mu}_{{\cal Z}_1{\cal Z}_2}^2\sim-\biggl[\left(\frac{1}{2}+\frac{{\cal O}({10}^3)}{2}\right)m_0^2-{(0.01)}(n^s)^2\hat{\mu}_{{\cal Z}_1{\cal Z}_2}^2+{(0.32)}m_{3/2}^2-1/2 M_{EW}^2+{(0.03)}n^s\hat{\mu}_{{\cal Z}_1{\cal Z}_2}m_{3/2}+\frac{19\pi}{2200}S_0\biggr].
\end{equation}
From (\ref{eq:light_H_II}) and (\ref{eq:muhat_II}) one therefore sees that the mass-squared of one of the two Higgs doublets, $m_{H_2}^2$, at the $EW$ scale is given by:
\begin{equation}
\label{eq:muhat_III}
\hskip -0.3in m^2_{H_2}=m^2_{{\cal Z}_2}+\hat{\mu}^2_{{\cal Z}_i{\cal Z}_i}= \left(\left(-\frac{1}{2}-\frac{{\cal O}({10}^3}{2}\right)m^2_{0}-{(0.06)}n^s\hat{\mu}_{{\cal Z}_1{\cal Z}_2}m_{3/2}\right)+\frac{1}{2}M_{EW}^2-\frac{19\pi}{1100}S_0.
\end{equation}
From \cite{D3_D7_Misra_Shukla}, we notice:
%\begin{equation}
%\label{eq:mu_m_Z_sq}
$\hat{\mu}_{{\cal Z}_1{\cal Z}_2}m_{3/2}\sim m_{{\cal Z}_i}^2,$
%\end{equation}
using which in (\ref{eq:muhat_III}), one sees that for an ${\cal O}(1)\ n^s$,
\begin{equation}
\label{eq:muhat_IV}
m^2_{H_2}(M_{EW})\sim\frac{1}{2}M_{EW}^2-\frac{19\pi}{1100}S_0-\frac{{\cal O}({10}^3)}{2}{\cal V}m_{3/2}^2.
\end{equation}
We have assumed at $m_{{\cal Z}_1}(M_s)=m_{{\cal Z}_2}(M_s)$. So, $S_0\approx m_{\rm squark/slepton}^2$, which in our setup could be of ${\cal O}(\hat{\mu}^2)$. Further,
\begin{equation}
\label{eq:heavy_Higgs}
m^2_{H_1}(M_{EW})=\left(m^2_{{\cal Z}_1}+\hat{\mu}_{{\cal Z}_1{\cal Z}_2}^2\right)(M_{EW})
 \sim m^2_{{\cal Z}_1}(M_s)+\frac{1}{2}M_{EW}^2+(0.01)(n^s)^2{\cal V}^2m_{3/2}^2.
\end{equation}
In the results on Soft SUSY Parameters summarized in Table 1, one finds that $\hat{\mu}B\sim\hat{\mu}^2$ at the string scale. By assuming the same to be valid at the string and EW scales, the Higgs mass matrix at the $EW$-scale can thus be expressed as:
\begin{eqnarray}
\label{eq:Higss_mass_matrix}
& & \left(\begin{array}{cc}
m^2_{H_1} & \hat{\mu}B\\
\hat{\mu}B & m^2_{H_2}\end{array}\right) \sim\left(\begin{array}{cc}
m^2_{H_1} & \xi\hat{\mu}^2\\
\xi\hat{\mu}^2 & m^2_{H_2}
\end{array}\right).
\end{eqnarray}
The eigenvalues are given by:
\begin{eqnarray}
\label{eq:eigenvalues}
& & \frac{1}{2}\biggl(m^2_{H_1}+m^2_{H_2}\pm\sqrt{\left(m^2_{H_1}-m^2_{H_1}\right)^2+4\xi^2\hat{\mu}^4}\biggr).
\end{eqnarray}
As (for ${\cal O}(1)\ n^s$)
\begin{eqnarray}
\label{eq:evs_1}
& & m^2_{H_1}+m^2_{H_2}\sim0.01{\cal V}^2m^2_{3/2}-0.06S_0+...,\nonumber\\
& & m^2_{H_1}-m^2_{H_2}\sim0.01{\cal V}^2m^2_{3/2}+0.06S_0+...,\nonumber\\
& & \hat{\mu}_{{\cal Z}_1{\cal Z}_2}^2\sim0.01{\cal V}^2m^2_{3/2}-0.03S_0+...,
\end{eqnarray}
one sees that the eigenvalues are:
\begin{eqnarray}
\label{eq:evs_2}
& & 0.01{\cal V}^2m^2_{3/2}-0.06S_0+...\pm\sqrt{\left(0.01{\cal V}^2m^2_{3/2}+0.06S_0+...\right)^2+\xi^2\left(0.02{\cal V}^2m^2_{3/2}-0.06S_0\right)^2}.\nonumber\\
& &
\end{eqnarray}
Hence, assuming a universality w.r.t. to the $D3$-brane position moduli masses ($m_{Z_{1,2}}$) and lack of the same for the squark/slepton masses, if $S_0$ and $\xi$ are fine tuned as follows:
\begin{equation}
\label{eq:evs_3}
0.01{\cal V}^2m^2_{3/2}\sim-0.06S_0\ {\rm and}\ \xi\sim\frac{2}{3}+\frac{{\cal O}(10)}{4{\cal V}^2}\left(\frac{m^2_{EW}}{m^2_{3/2}}\right),
\end{equation}
one sees that one obtains one light Higgs doublet (corresponding to the negative sign of the square root) with a mass of about $125 GeV$
%which could be tuned to the $EW$ scale if $\hat{\mu}^2\sim m^2_{H_i}+m_{EW}^2$
and one heavy Higgs doublet (corresponding to the positive sign of the square root). Note, however, the Higgsino mass parameter $\hat{\mu}_{{\cal Z}_1{\cal Z}_2}$ then turns out to be heavy with a value, at the EW scale of around $0.01{\cal V}m_{3/2}$  i.e to the order of squark/slepton mass squared scale which is possible in case of $\mu$ split SUSY scenario discussed above.  This shows the possibility of realizing  $\mu$ split SUSY scenario in the context of LVS phenomenology named as large volume ``$\mu$-split SUSY" scenario.

\subsection{Obtaining Long-Lived Gluinos}

The most distinctive feature of split SUSY, decisively differentiating it from the usual Supersymmetric Standard Model, is based on longevity of the gluinos. Since the squarks which mediate its decay are extremely heavy, one expects life time of Gluinos to be high. The  decay amplitudes for the three-body tree-level and two-body one-loop diagrams of Fig.1 were evaluated in \cite{Dhuria+Misra_mu_Split SUSY} by considering the contribution of relevant terms in gauged supergravity action of Wess and Bagger \cite{Wess_Bagger} given below:
\begin{eqnarray}
 & &  {\cal L}  =  g_{YM}g_{\sigma_B {\bar J}}X^{\sigma_B}{\bar\chi}^{\bar J}\lambda_{\tilde{g}} {\hskip -0.05in}+ i \sqrt{g}g_{I{\bar j}}{\bar\chi}^{\bar j}{\bar\sigma}^\mu\bigtriangledown_\mu\chi^I\nonumber\\
 & & + \frac{e^{\frac{K}{2}}}{2}\left({\cal D}_iD_JW\right)\chi^i\chi^J {\hskip -0.05in}
  +g_{YM}g_{I{\bar J}}{\bar\chi}^{\bar J}{\bar\sigma}\cdot A\ {\rm Im}\left(X^{\sigma_B}K + i D^{\sigma_B}\right) {\hskip -0.05in}\chi^I;\nonumber\\
& &
 \end{eqnarray}
 $W$ is the superpotential as defined in \cite{D3_D7_Misra_Shukla,ferm_masses_MS,Dhuria+Misra_mu_Split SUSY}, $\sigma_B$ is the complexified (by four-form axions) big divisor volume, $\chi/{\bar\chi},\lambda_{\tilde{g}}$ correspond to quarks/antiquarks and gaugino's and $X^{\sigma_B}=-6i\kappa_4^2\mu_7Q_B$, where $Q_B=2\pi\alpha^\prime\int_{\Sigma_B}i^*\omega_\alpha\wedge P_-\tilde{f}$ where $P_-$ is a harmonic zero-form on $\Sigma_B$ taking value +1 on $\Sigma_B$ and $-1$ on $\sigma(\Sigma_B)$ - $\sigma$ being a holomorphic isometric involution as part of the Calabi-Yau orientifold - and $\tilde{f}\in\tilde{H}^2_-(\Sigma^B)\equiv{\rm coker}\left(H^2_-(CY_3)\stackrel{i^*}{\rightarrow}H^2_-(\Sigma^B)\right)$; $D^{\sigma_B}=\frac{4\pi\alpha^\prime\kappa_4^2\mu_7Q_Bv^B}{\cal V}$. It is understood that the $D3$-brane position moduli are indexed by $i$ and the Wilson line moduli are indexed by $I=1,..,h^{0,1}_-(\Sigma_B)$.

   \begin{center}
\begin{picture}(500,150)(50,0)
%%%%%%%%%%%%%%%%%%%%%%%%%%%%%%
\Text(90,130)[]{$\tilde{g}$}
%\ArrowLine(110,120)(160,120)
\Line(60,120)(100,120)
\Gluon(60,120)(100,120){5}{4}
\ArrowLine (125,150)(100,120)
\Text(127,153)[]{$\bar{q}_I$}
\DashArrowLine (100,120)(120,90){4}
\Text(104,100)[]{$\tilde{q}_J$}
\ArrowLine(120,90)(145,120)
\Text(152,123)[]{$\tilde{\chi}_3^0$ }
\ArrowLine(120,90)(145,60)
\Text(150,66)[]{${q_K}$}

%%%%%%%%%%%%%%%%%%%%%%
\Text(195,130)[]{$\tilde{g}$}
\Line(165,120)(205,120)
\Gluon(165,120)(205,120){5}{4}
\ArrowLine (230,150)(205,120)
\Text(234,153)[]{$q_K$}
\DashArrowLine(205,120)(225,90){4}{}
\Text(208,100)[]{$\tilde{q}_J$}
\ArrowLine(225,90)(250,120)
\Text(256,122)[]{$\bar{q}_I$}
\ArrowLine(225,90)(250,60)
\Text(256,66)[]{$\tilde{\chi}_3^0$}

%%%%%%%%%%%%%%%%%%%%%%%%%

%%%%%%%%%%%%%%%%%%%%%%%%%%%%%%

\Line(270,120)(310,120)
\Gluon(270,120)(310,120){5}{4}
\Text(300,130)[]{$\tilde{g}$}
\ArrowLine (310,120)(335,150)
\Text(320,140)[]{$q_I$}
\DashArrowLine (335,70)(310,120){4}
\Text(318,90)[]{$\tilde{q}_R$}
\DashArrowLine (335,150)(335,70){4}
\Gluon(335,70)(365,70){4}{3}
\Text(373,68)[]{$g_\mu$}
\ArrowLine(365,150)(335,150)
\Text(374,154)[]{$\tilde{\chi}_3^0$}

%%%%%%%%%%%%%%%%%%%%%%%%%%%%

%\ArrowLine(110,120)(160,120)
\Line(380,120)(420,120)
\Gluon(380,120)(420,120){5}{4}
\Text(300,130)[]{$\tilde{g}$}
\DashArrowLine (420,120)(445,150){4}
\Text(422,134)[]{$\tilde{q}_R$}
\ArrowLine (445,70)(420,120)
\Text(422,100)[]{$q_I$}
\ArrowLine (445,150)(445,70)
\Gluon(445,70)(475,70){4}{3}
\Text(483,68)[]{$g_\mu$}
\ArrowLine(475,150)(445,150)
\Text(484,151)[]{$\tilde{\chi}_3^0$}

\end{picture}
\vskip -0.7in

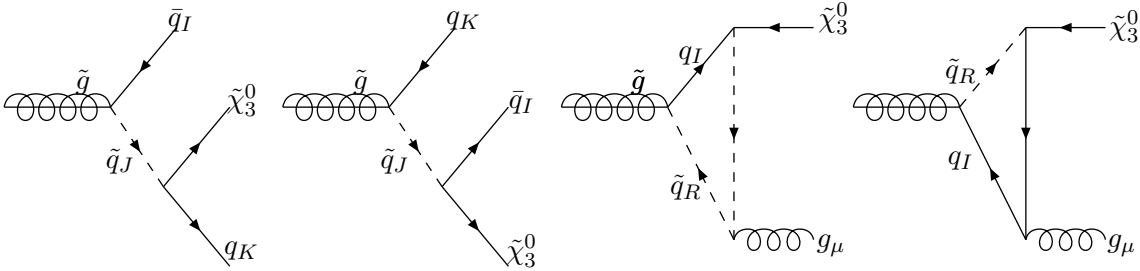
{\sl Fig. 1: Decay channels corresponding to tree level three body as well as one loop two body decay of Gluino}
\end{center}

For the two Wilson line moduli set up summarized in {\bf 1}, by  identifying the fermionic superpartners of the Wilson line moduli $a_1, a_2$'s with (anti-)quarks, the contribution of tree level as well as one-loop Feynman diagrams can be worked out from the following vertices:
 $${\hskip -2.2in} {\rm Gluino-quark-squark}~  {\tilde G}^{{q/{\bar q}}}_{\tilde{q}}: \tilde{f}\left({\cal V}^{-\frac{37}{36}}\delta^I_{a_1}+{\cal V}^{-\frac{59}{36}}\delta^I_{a_2}\right)$$
$${\hskip -0.3in} {\rm Neutralino-quark-squark }~X^{{q}}_{\tilde{q}}{\hskip -0.05in}:i\tilde{f}\left[\left({\cal V}^{-\frac{31}{72}}\frac{{\bar\sigma}\cdot p_{\tilde{\chi}_3^0}}{M_p}+ {\cal V}^{-\frac{37}{36}}\right)\delta^I_{a_1}+ \left({\cal V}^{-\frac{67}{72}}\frac{{\bar\sigma}\cdot p_{\tilde{\chi}_3^0}}{M_p} + {\cal V}^{-\frac{59}{36}}\right)\delta^I_{a_2}\right]  $$
 $${\hskip -2.0in}{\rm Gluon-squark-squark}~{ G}^{\tilde q}_{\tilde{q}}:{{\cal V}^{-\frac{4}{3}}{\tilde f}\left[ 2\epsilon\cdot k-\epsilon\cdot\left(p_{\tilde{\chi}_3^0}+p_{\tilde{g}}\right)\right]}$$
$${\hskip -0.9in}{\rm Gluon-quark-quark }~{ G}^{{q/{\bar q}}}_{{q/{\bar q}}}:\left({\cal V}^{-\frac{3}{4}}\delta^I_{a_1}\delta^J_{a_1}+{\cal V}^{-\frac{5}{4}}\delta^I_{a_{1/2}}\delta^J_{a_{2/1}}+{\cal V}^{-\frac{7}{4}}\delta^I_{a_2}\delta^J_{a_2}\right)\tilde{f}{\bar\sigma}\cdot\epsilon. $$ In the same, the neutralino is defined as (See \cite{Dhuria+Misra_mu_Split SUSY}):
$\tilde{\chi}_3^0\sim-\lambda^0+\tilde{f}{\cal V}^{-\frac{51}{72}}\left(\tilde{H}_1^0+\tilde{H}_2^0\right);\ {\rm mass}\sim
\frac{1}{2}{\cal V}^{-2}M_p<m_{\frac{3}{2}}$ and $\tilde{H}_{1,2}^0$ are the Higgsinos.

The upper bound on ${\tilde f}$ evaluated in dilute flux approximation by demanding the flux-generated D-term potential is sub-dominant as compared to the F-term potential (See \cite{Dhuria+Misra_mu_Split SUSY}), is ${10}^{-4}$. Using RG
analysis of coefficients of the effective dimension-six gluino decay operators as given in
\cite{Guidice_et_al}, it was shown in \cite{Dhuria+Misra_mu_Split SUSY} that these coefficients at the EW scale are of the same order as that at the squark mass scale.

 By considering four different possibilities of identifying quarks/antiquarks with the fermionic superpartners of the Wilson line moduli $a_{1,2}$, the lower bound on the gluino lifetime via this three-body decay channel come out to be : {$10^{-9}\delta^I_1\delta^J_1\delta^K_1{\rm sec}, 10^{-2}\delta^I_2\delta^J_1\delta^K_1{\rm sec}, 10^{8}\delta^I_2\delta^J_1\delta^K_2  {\rm sec}, 10^{-2}\delta^I_1\delta^J_1\delta^K_2{\rm sec}$} for tree level Gluino decay and ${10}^4{\rm sec}$ for one-loop two-body gluino decay. Adopting the same approach,  we  calculated in \cite{Dhuria+Misra_mu_Split SUSY} the decay width of tree-level as well as one-loop two-body gluino decays into Goldstino, results of which, similar to the tree level gluino decay into neutralino, yield large life time(s) of gluino.

\section{Obtaining Ricci-Flat Swiss-Cheese Metrics in the Large Volume}

In principle, due to the presence of a mobile $D3$-brane, one must also include the geometric K\"{a}hler potential $K_{\rm geom}$ of the Swiss-Cheese Calabi-Yau in the moduli space K\"{a}hler potential. In \cite{D3_D7_Misra_Shukla}, given that we had restricted the mobile $D3$-brane to $\Sigma_B$, one had estimated (in the large volume limit) $K_{\rm geom}\sim
\frac{{\cal V}^{-\frac{1}{3}}}{\sqrt{ln {\cal V}}}$ summarized as follows. Using GLSM techniques and the toric data for the given Swiss-Cheese Calabi-Yau, the geometric K\"{a}hler potential for the divisor ${\Sigma_B}$ (and ${\Sigma_S}$) in the LVS limit was evaluated in \cite{D3_D7_Misra_Shukla} in terms of derivatives of genus-two Siegel theta functions as well as
two Fayet-Iliopoulos parameters corresponding to the two $C^*$ actions in the  two-dimensional ${\cal N}=2$ supersymmetric gauge theory whose target space is our toric variety Calabi-Yau, and a parameter $\zeta$ encoding the information about the $D3-$brane position moduli-independent (in the LVS limit) period matrix of the hyperelliptic curve $w^2=P(z)$, $P(z)$ being the sextic in the exponential of the vector superfields eliminated as auxiliary fields, corresponding to $\Sigma_B$. To be a bit more specific, one can show that upon elimination of the vector superfield (in the IR limit of the GLSM), one obtains an octic in $e^{2V_2}$, $V_2$ being one of the two real gauge superfields. Using Umemura's result \cite{Umemura} on expressing the roots of an algebraic polynomial of  degree $n$ in terms of Siegel theta functions
of genus $g(>1)=[(n+2)/2]$ :  $\theta\left[\begin{array}{c} \mu\\
\nu
\end{array}\right](z,\Omega)$ for $\mu,\nu\in{\bf R}^g, z\in {\bf C}^g$ and $\Omega$ being a complex symmetric
$g\times g$ period matrix with $Im(\Omega)>0$ defined as follows:
$$
\theta\left[\begin{array}{c} \mu\\
\nu
\end{array}\right](z,\Omega)=\sum_{n\in{\bf Z}^g}e^{i\pi(n+\mu)^T\Omega(n+\mu)+2i\pi(n+\mu)^T(z+\nu)}.$$
Hence for an octic, one needs to use Siegel theta functions of genus five. The period matrix $\Omega$ will be defined as follows:
$$\Omega_{ij}=\left(\sigma\right)^{-1}_{ik}\rho_{kj}$$
where $$\sigma_{ij}\equiv\oint_{A_j}dz \frac{z^{i-1}}{\sqrt{z(z-1)(z-2)P(z)}}$$ and
$$\rho_{ij}\equiv\oint_{B_j}\frac{z^{i-1}}{\sqrt{z(z-1)(z-2)P(z)}},$$
$\{A_i\}$ and $\{B_i\}$ being a canonical basis of cycles satisfying: $A_i\cdot A_j=B_i\cdot B_j=0$ and
$A_i\cdot B_j=\delta_{ij}$. Umemura's result then is that a root:
$$\hskip-0.3in\frac{1}{2\left(\theta\left[\begin{array}{ccccc}
\frac{1}{2} & 0 & 0 & 0 & 0 \\
0 & 0 & 0 & 0 & 0  \end{array}\right](0,\Omega)\right)^4
\left(\theta\left[\begin{array}{ccccc}
\frac{1}{2} & \frac{1}{2} & 0 & 0 & 0 \\
0 & 0 & 0 & 0 & 0  \end{array}\right](0,\Omega)\right)^4}$$
$$\hskip-0.3in\times\Biggl[\left(\theta\left[\begin{array}{ccccc}
\frac{1}{2} & 0 & 0 & 0 & 0 \\
0 & 0 & 0 & 0 & 0  \end{array}\right](0,\Omega)\right)^4\left(\theta\left[\begin{array}{ccccc}
\frac{1}{2} & \frac{1}{2} & 0 & 0 & 0 \\
0 & 0 & 0 & 0 & 0  \end{array}\right](0,\Omega)\right)^4$$
$$\hskip-0.3in+ \left(\theta\left[\begin{array}{ccccc}
0 & 0 & 0 & 0 & 0 \\
0 & 0 & 0 & 0 & 0  \end{array}\right](0,\Omega)\right)^4
\left(\theta\left[\begin{array}{ccccc}
0 & \frac{1}{2} &  0 & 0 & 0 \\
0 & 0 & 0 & 0 & 0  \end{array}\right](0,\Omega)\right)^4$$
$$\hskip-0.3in- \left(\theta\left[\begin{array}{ccccc}
0 & 0 & 0 & 0 & 0 \\
\frac{1}{2} & 0 & 0 & 0 & 0  \end{array}\right](0,\Omega)\right)^4
\left(\theta\left[\begin{array}{ccccc}
0 & \frac{1}{2} & 0 & 0 & 0 \\
\frac{1}{2} & 0 & 0 & 0 & 0 \end{array} \right](0,\Omega)\right)^4\Biggr].$$
In the LVS limit, the octic reduces to a sextic. Umemura's result would require the use of genus-four Siegel theta functions. However, using the results of
\cite{Zhivkov}, one can express the roots of a sextic in terms of derivatives of genus-two Siegel theta functions as follows:
$$\hskip-0.6in\left[\frac{\sigma_{22}\frac{d}{dz_1}\theta\left[\begin{array}{cc}
\frac{1}{2}&\frac{1}{2} \\
0&\frac{1}{2}
\end{array}\right]\left((z_1,z_2),\Omega\right)
- \sigma_{21}\frac{d}{dz_2}\theta\left[\begin{array}{cc}
\frac{1}{2}&\frac{1}{2} \\
0&\frac{1}{2}
\end{array}\right]\left((z_1,z_2),\Omega\right) }
{\sigma_{12}\frac{d}{dz_1}\theta\left[\begin{array}{cc}
\frac{1}{2}&\frac{1}{2} \\
0&\frac{1}{2}
\end{array}\right]\left((z_1,z_2),\Omega\right)
- \sigma_{12}\frac{d}{dz_2}\theta\left[\begin{array}{cc}
\frac{1}{2}&\frac{1}{2} \\
0&\frac{1}{2}
\end{array}\right]\left((z_1,z_2),\Omega\right)}\right]_{z_1=z_2=0},$$
etc. The symmetric period matrix corresponding to the hyperelliptic
curve $w^2=P(z)$ is given by:
$$\hskip-0.3in\left(\begin{array}{cc}
\Omega_{11} & \Omega_{12} \\
\Omega_{12} & \Omega_{22}
\end{array}\right)=\frac{1}{\sigma_{11}\sigma_{22}-\sigma_{12}\sigma_{21}}\left(\begin{array}{cc}
\sigma_{22} & -\sigma_{12} \\
-\sigma_{21} & \sigma_{11}
\end{array}\right)\left(\begin{array}{cc}
\rho_{11} & \rho_{12} \\
\rho_{21} & \rho_{22}
\end{array}\right),$$
where $\sigma_{ij}=\int_{z_*{A_j}}\frac{z^{i-1}dz}{\sqrt{P(z)}}$ and
$\rho_{ij}=\int_{z_*{B_j}}\frac{z^{i-1}dz}{\sqrt{P(z)}}$ where $z$ maps the $A_i$ and $B_j$ cycles to the
$z-$plane. The geometric K\"{a}hler potential for the divisor $\Sigma_B$ in the LVS limit, as shown in \cite{D3_D7_Misra_Shukla}, then turns out to be  given by:
\vskip-0.1in
\begin{eqnarray}
\label{eq:Kaehler_D_5}
 K|_{\Sigma_B} & \sim &   r_2 - {\left[r_2 - \left(1+|z_1|^2+|z_2|^2\right)\left(\frac{\zeta}{
r_1|z_3|^2}\right)^{\frac{1}{6}}\right]}{4\sqrt{\zeta}}/{3}\nonumber\\
 & & +|z_3|^2\left\{{\left[r_2 - \left(1+|z_1|^2+|z_2|^2\right)\left(\frac{\zeta}{
r_1|z_3|^2}\right)^{\frac{1}{6}}\right]}{\sqrt{\zeta}}/{3\sqrt{r_1|z_3|^2}}\right\}^2\nonumber\\
& & - r_1 ln\left\{{\left[r_2 - \left(1+|z_1|^2+|z_2|^2\right)\left({\zeta}{
r_1|z_3|^2}\right)^{\frac{1}{6}}\right]}\sqrt{\zeta}/{3\sqrt{r_1|z_3|^2}}\right\}
-r_2 ln\left[\left({\zeta}/{r_1|z_3|^2}\right)^{\frac{1}{6}}\right]\nonumber\\
 & & \sim{{\cal V}^{\frac{2}{3}}}/{\sqrt{ln {\cal V}}}.
\end{eqnarray}
\vskip-0.13in
\noindent As mentioned earlier, if the space-time filling mobile $D3$-brane is free to explore the full Calabi-Yau, one would require the knowledge of the geometric K\"{a}hler potential of the full Calabi-Yau. We will now estimate $K_{\rm geom}$ using the Donaldson's algorithm \cite{Donaldson_i} and obtain a metric for the Swiss-Cheese Calabi-Yau at a generic point finitely separated from $\Sigma_B$, that is Ricci-flat in the large volume limit; we should note that GLSM-based metrics are not expected to yield Ricci-flat metrics.

The crux of the Donaldson's algorithm is that the sequence
$\frac{1}{k\pi}\partial_i{\bar\partial}_{\bar j}\left(ln\sum_{\alpha,\beta}h^{\alpha{\bar\beta}}s_\alpha{\bar s}_{\bar\beta}\right)$
on $P(\{z_i\})$, in the $k\rightarrow\infty$-limit - which in practice implies $k\sim10$ - converges to a unique Calabi-Yau metric for the given K\"{a}hler class and complex structure; $h_{\alpha{\bar\beta}}$ is a balanced metric on the line bundle ${\cal O}_{P(\{z_i\})}(k)$ (with sections $s_\alpha$) for any $k\geq1$, i.e.,
\vskip-0.2in
\begin{equation}
\label{balanced_metric}
T(h)_{\alpha{\bar\beta}}\equiv \frac{N_k}{\sum_{j=1}w_j}\sum_{i}\frac{s_\alpha(p_i)\overline{s_\beta(p_i)}w_i}
{h^{\gamma{\bar\delta}}
s_\gamma(p_i)\overline{s_\delta(p_i)}}=h_{\alpha{\bar\beta}},
\end{equation}
%\vskip-0.2in
\noindent where the weight at point $p_i$, $w_i\sim\frac{i^*(J_{GLSM}^3)}{\Omega\wedge{\bar\Omega}}$ with the embedding map $i:P(\{z_i\})\hookrightarrow{\bf WCP}^4$ and the number of sections is denoted by $N_k$.  The defining hypersurface of the Swiss-Cheese Calabi-Yau in the $x_2=1$-coordinate patch in ${\bf WCP}^4[1,1,1,6,9]$ is given by:
%\begin{equation}
%\label{eq:hyp_def}
$1+z_1^{18}+z_2^{18}+z_3^3+z_4^2-\psi z_1z_2z_3z_4-3\phi z_1^6z_2^6=0.$
%\end{equation}
In the large volume limit, the above can be satisfied if, e.g., $1+z_1^{18}+z_2^{18}\sim-z_3^3$,
$z_4^2\sim\psi z_1z_2z_3z_4+3\phi z_1^6z_2^6$. For $z_{1,2}\sim{\cal V}^{\frac{1}{36}}$, one sees the same are satisfied for $z_{3,4}\sim {\cal V}^{\frac{1}{6}}$ provided $\psi {\cal V}^{\frac{1}{18}}\sim 3\phi$. Therefore:
\begin{eqnarray}
\label{eq:sec_metr_h}
& & h_{1{\bar z}_i}\sim\frac{{\cal V}^{\frac{1}{36}}}{h^{z_4^2{\bar z}_4^2}{\cal V}^{\frac{2}{3}}},\  h_{1{\bar z}_4}\sim\frac{{\cal V}^{\frac{1}{6}}}{h^{z_4^2{\bar z}_4^2}{\cal V}^{\frac{2}{3}}}\nonumber\\
& & h_{1{\bar z}_i^2}\sim\frac{{\cal V}^{\frac{1}{18}}}{h^{z_4^2{\bar z}_4^2}{\cal V}^{\frac{2}{3}}},\  h_{1{\bar z}_4^2}\sim\frac{{\cal V}^{\frac{1}{3}}}{h^{z_4^2{\bar z}_4^2}{\cal V}^{\frac{2}{3}}}\nonumber\\
& & h_{1{\bar z}_i{\bar z}_4}\sim\frac{{\cal V}^{\frac{1}{6}+\frac{1}{36}}}{h^{z_4^2{\bar z}_4^2}{\cal V}^{\frac{2}{3}}},\  etc.,
\end{eqnarray}
which on being inverted gives:
\begin{eqnarray}
\label{eq:metsecinv}
& h^{\alpha{\bar\beta}}\sim &  \left(
\begin{array}{llllll}
 h^{z_4^2{\bar z}_4^2} V^{2/3} & h^{z_4^2{\bar z}_4^2} V^{23/36} & h^{z_4^2{\bar z}_4^2} \sqrt{V} & h^{z_4^2{\bar z}_4^2} V^{11/18} & h^{z_4^2{\bar z}_4^2} \sqrt[3]{V} & h^{z_4^2{\bar z}_4^2} V^{17/36} \\
 h^{z_4^2{\bar z}_4^2} V^{23/36} & h^{z_4^2{\bar z}_4^2} V^{11/18} & h^{z_4^2{\bar z}_4^2} V^{17/36} & h^{z_4^2{\bar z}_4^2} V^{7/12} & h^{z_4^2{\bar z}_4^2} V^{11/36} & h^{z_4^2{\bar z}_4^2} V^{4/9} \\
 h^{z_4^2{\bar z}_4^2} \sqrt{V} & h^{z_4^2{\bar z}_4^2} V^{17/36} & h^{z_4^2{\bar z}_4^2} \sqrt[3]{V} & h^{z_4^2{\bar z}_4^2} V^{4/9} & h^{z_4^2{\bar z}_4^2} \sqrt[6]{V} & h^{z_4^2{\bar z}_4^2} V^{11/36} \\
 h^{z_4^2{\bar z}_4^2} V^{11/18} & h^{z_4^2{\bar z}_4^2} V^{7/12} & h^{z_4^2{\bar z}_4^2} V^{4/9} & h^{z_4^2{\bar z}_4^2} V^{5/9} & h^{z_4^2{\bar z}_4^2} V^{5/18} & h^{z_4^2{\bar z}_4^2} V^{5/12} \\
 h^{z_4^2{\bar z}_4^2} \sqrt[3]{V} & h^{z_4^2{\bar z}_4^2} V^{11/36} & h^{z_4^2{\bar z}_4^2} \sqrt[6]{V} & h^{z_4^2{\bar z}_4^2} V^{5/18} & h^{z_4^2{\bar z}_4^2} & h^{z_4^2{\bar z}_4^2} V^{5/36} \\
 h^{z_4^2{\bar z}_4^2} V^{17/36} & h^{z_4^2{\bar z}_4^2} V^{4/9} & h^{z_4^2{\bar z}_4^2} V^{11/36} & h^{z_4^2{\bar z}_4^2} V^{5/12} & h^{z_4^2{\bar z}_4^2} V^{5/36} & h^{z_4^2{\bar z}_4^2} V^{5/18}
\end{array}
\right).\nonumber\\
& &
\end{eqnarray}
Using (\ref{eq:metsecinv}), one hence obtains the following K\"{a}hler potential ansatz:
\begin{eqnarray}
\label{eq:K}
& & \hskip-0.5in K=ln \Biggl[h^{z_4^2{\bar z}_4^2} {z_4}^2 {{\bar z}_4}^2+h^{z_4^2{\bar z}_4^2} \sqrt[3]{V} {z_4} {{\bar z}_4}+h^{z_4^2{\bar z}_4^2} V^{23/36}
   ({z_1}+{{\bar z}_1}+{z_2}+{{\bar z}_2})+h^{z_4^2{\bar z}_4^2} V^{11/18} ({z_1} {{\bar z}_1}+{z_2} {{\bar z}_1}+{z_1}
   {{\bar z}_2}+{z_2} {{\bar z}_2})\nonumber\\
   & &\hskip-0.5in +h^{z_4^2{\bar z}_4^2} V^{11/18} \left({z_1}^2+{z_2}
   {z_1}+{{\bar z}_1}^2+{z_2}^2+{{\bar z}_2}^2+{{\bar z}_1} {{\bar z}_2}\right) +h^{z_4^2{\bar z}_4^2} V^{7/12} \left({{\bar z}_2}
   {z_1}^2+{{\bar z}_1}^2 {z_1}+{{\bar z}_2}^2 {z_1}+{{\bar z}_1} {z_2}^2+{z_2} {{\bar z}_2}^2+{{\bar z}_1}^2
   {z_2}\right)\nonumber\\
   & &\hskip-0.5in +h^{z_4^2{\bar z}_4^2} V^{5/9} \left({z_1}^2 {{\bar z}_1}^2+ {z_2}^2 {{\bar z}_1}^2+ {z_1}^2
   {{\bar z}_2}^2+{z_2}^2 {{\bar z}_2}^2\right)+h^{z_4^2{\bar z}_4^2} \sqrt{V} ({z_4}+{{\bar z}_4})+ h^{z_4^2{\bar z}_4^2} V^{17/36} ({{\bar z}_1}
   {z_4}+{{\bar z}_2} {z_4}+{z_1} {{\bar z}_4}+{z_2} {{\bar z}_4})\nonumber\\
   & & \hskip-0.5in +h^{z_4^2{\bar z}_4^2} V^{4/9} \left({{\bar z}_4}
   {z_1}^2+{{\bar z}_1}^2 {z_4}+{{\bar z}_2}^2 {z_4}+{z_2}^2 {{\bar z}_4}\right)+h^{z_4^2{\bar z}_4^2} V^{5/12} \left({{\bar z}_1}
   {{\bar z}_4} {z_1}^2+{{\bar z}_2} {{\bar z}_4} {z_1}^2+{{\bar z}_1}^2 {z_4} {z_1}+{z_2} {{\bar z}_2}^2
   {z_4}+{{\bar z}_1} {z_2}^2 {{\bar z}_4}+{z_2}^2 {{\bar z}_2} {{\bar z}_4}\right)\nonumber\\
   & &\hskip-0.5in +h^{z_4^2{\bar z}_4^2} V^{5/18} ({z_1}
   {{\bar z}_1} {z_4} {{\bar z}_4}+ {{\bar z}_1} {z_2} {z_4} {{\bar z}_4}+ {z_1} {{\bar z}_2} {z_4}
   {{\bar z}_4}+{z_2} {{\bar z}_2} {z_4} {{\bar z}_4})+h^{z_4^2{\bar z}_4^2} V^{11/36} (({z_1}+{{\bar z}_1}) {z_4}
   {{\bar z}_4}+({z_2}+{{\bar z}_2}) {z_4} {{\bar z}_4})\nonumber\\
   & & \hskip-0.5in +h^{z_4^2{\bar z}_4^2} \sqrt[3]{V} \left({z_4}^2+{{\bar z}_4}^2\right)+h^{z_4^2{\bar z}_4^2}
   V^{11/36} \left({z_4}^2 {\bar z}_1+{z_1} {{\bar z}_4}^2+{z_2} {{\bar z}_4}^2+{\bar z}_2 {z_4}^2\right)+h^{z_4^2{\bar z}_4^2}
   V^{5/18} \left({{\bar z}_1}^2 {z_4}^2+{{\bar z}_2}^2 {z_4}^2+{z_1}^2 {{\bar z}_4}^2+{z_2}^2
   {{\bar z}_4}^2\right)\nonumber\\
   & & \hskip-0.5in +h^{z_4^2{\bar z}_4^2} \sqrt[6]{V} \left({{\bar z}_4} {z_4}^2+{{\bar z}_4}^2 {z_4}\right)+h^{z_4^2{\bar z}_4^2} V^{5/36}
   \left(({{\bar z}_1}+{{\bar z}_2}) {{\bar z}_4} {z_4}^2+({z_1}+{z_2}) {{\bar z}_4}^2 {z_4}\right)+h^{z_4^2{\bar z}_4^2} V^{4/9}
   ({z_1} {{\bar z}_2} {z_4}+{{\bar z}_1} {z_2} {{\bar z}_4}+{z_1} {{\bar z}_1}
   ({z_4}+{{\bar z}_4})\nonumber\\
   & & \hskip-0.5in+{z_1} {{\bar z}_2} ({z_4}+{{\bar z}_4})+{z_2} {{\bar z}_2}
   ({z_4}+{{\bar z}_4}))+\sqrt[3]{V}\Biggr]
\end{eqnarray}
From GLSM-based analysis, we had seen in (\ref{eq:Kaehler_D_5}) that on $\Sigma_B(z_4=0)$, the argument of the logarithm received the most dominant contribution from the FI-parameter $r_2\sim{\cal V}^{\frac{1}{3}}$. From (\ref{eq:metsecinv}), one sees that $h^{1{\bar 1}}\sim h^{z_4^2{\bar z}_4^2}{\cal V}^{\frac{2}{3}}$. For consistency, we should therefore obtain $h^{z_4^2{\bar z}_4^2}\sim {\cal V}^{-\frac{1}{3}}$. This has been assumed in (\ref{eq:K}) and will be verified below to correspond to one allowed value of $h^{z_4^2{\bar z}_4^2}$ that would yield an approximately Ricci-flat metric (up to within 10$\%$). Using (\ref{eq:K}), one can show that:
\begin{equation}
\label{eq:R11bar_i}
R_{z_i{\bar z}_j}\sim\frac{\sum_{n=0}^8a_n\left(h^{z_4^2{\bar z}_4^2}\right)^n {\cal V}^{\frac{n}{3}}}{\left(1+{\cal O}(1)h^{z_4^2{\bar z}_4^2} {\cal V}^{\frac{1}{3}}\right)^2{\cal V}^{\frac{1}{18}}\left(\sum_{n=0}^3b_n\left(h^{z_4^2{\bar z}_4^2}\right)^n {\cal V}^{\frac{n}{3}}\right)^2}.
\end{equation}
Solving numerically: $\sum_{n=0}^8a_n\left(h^{z_4^2{\bar z}_4^2}\right)^n {\cal V}^{\frac{n}{3}}=0$, as was assumed, one (of the eight values of) $h^{z_4^2{\bar z}_4^2}$, up to a trivial K\"{a}hler transformation, turns out to be ${\cal V}^{-\frac{1}{3}}, {\cal V}\sim 10^6$!! Using this value of $h^{z_4^2{\bar z}_4^2}$,
one obtains: $R_{z_i{\bar z}_4},R_{z_4{\bar z}_4}\sim10^{-1}$. Further, as has been assumed that the metric components $g_{z_{1,2}{\bar z}_4}$ are negligible as compared to $g_{z_i{\bar z}_j}$ - this was used in showing the completeness of the basis spanning $H^{1,1}_-$ in the large volume limit - is born out explicitly, wherein the latter turn out to about 10$\%$ of the former.

\section{Conclusion and Discussion}

 We have reviewed recent progress in realizing  $\mu$-split SUSY scenario localized around a mobile space-time filling $D3$-brane in the  context of type IIB Swiss-Cheese orientifold (involving isometric holomorphic involution) compactifications
 in the L(arge) V(olume) S(cenarios).
 Generation of very heavy scalars and light(superpartner) fermions
that had already been obtained in the context of L(arge) V(olume) S(cenario) in \cite{D3_D7_Misra_Shukla,ferm_masses_MS} and reviewed in {\bf 1} and {\bf 2.1}, was adopted as one of the signatures of split supersymmetric behavior. To see it more clearly, in \cite{Dhuria+Misra_mu_Split SUSY} and reviewed in {\bf 2.1}, we showed how to generate one
 light Higgs boson with the assumption  that fine tuning is allowed in case of split SUSY models. For this, using solution of  RG flow equation for the mobile $D3$-brane position moduli masses and Higgsino mass term and    further assuming gauge coupling up to one loop order and non-universality in squark/slepton masses (in addition to the non-universality between the Higgs' and squark/slepton masses), by diagonalizing the mass
   matrix for the Higgs doublet, we showed how  one could obtain one light Higgs (about $125 GeV$)   and one heavy Higgs, about a tenth of the squark masses. The  Higgsino also turns out to about a tenth of the squark mass. Since in our setup, $\mu$ value comes out to be of the order of squark/slepton mass scale i.e high scale, therefore we see the possibility of explicitly realizing $\mu$-split SUSY scenario which we could refer to as ``Local  $D3/D7\ \mu$-split SUSY Scenario".

The most distinctive feature of split SUSY is based on longevity of gluino. Therefore, in order to seek striking evidence of split SUSY in the context of LVS, in \cite{Dhuria+Misra_mu_Split SUSY} and reviewed in {\bf 2.2},  we  estimated the decay width for tree-level three-body gluino decay
into a quark, squark and neutralino. By constructing the neutralino mass matrix and diagonalizing
the same, we had identified the neutralino with a mass less than that of the gluino (this neutralino in the dilute flux approximation
is roughly half the mass of the gluino).  This neutralino turns out to be largely a neutral gaugino with a small admixture of the Higgsinos. Using one-loop RG analysis of coefficients of the effective dimension-six gluino decay operators as given in \cite{Guidice_et_al}, we had showed in \cite{Dhuria+Misra_mu_Split SUSY} that these coefficients at the EW scale are of the same order as that at the squark mass scale; we assume that these coefficients at the EW scale will be of the same order as that at the string scale. The lower bound on the gluino lifetime via this three-body decay channel was estimated to lie in the range: $10^{-9}-10^{6}$ seconds depending on which  two Wilson line moduli are used to model the two (anti-)quarks produced in gluino decay. We had also calculated the decay width of one-loop
two-body gluino decay into gluon and neutralino in \cite{Dhuria+Misra_mu_Split SUSY}, results of which, similar to the tree level gluino decay, yield large life time(s) of gluino  for this case. The high squark mass, helps to suppress the tree-level as well as one loop gluino decay width. The fact that we have obtained  suppressed  Gluino decay width  for squark masses of the order of $10^{12}GeV$, is in agreement with the previous theoretical studies based on gluino decays in split SUSY in literature (\cite{Guidice_et_al,Manuel Toharia}) and results based on collider phenomenology for stable gluino.

We are currently looking into exploring the neutralino to be a dark matter candidate in  a four-Wilson-line moduli (corresponding to the $SU(2)_L$ first-generation quark doublet, $SU(2)_L$ first-generation quark singlet, $SU(2)_L$ first-generation lepton doublet and the first generation $SU(2)_L$ first-generation lepton singlet) local $D3/D7$ $\mu$-Split SUSY framework \cite{neutralino_DM}.

\section*{Acknowledgments}

   We would like to thank P.Shukla and M.Dhuria for enjoyable collaborations on the issues reviewed in this article; we also thank M.Dhuria for helping out in drawing the Feynman diagrams. We would also thank the Abdus Salam ICTP (under the regular associateship scheme), CERN, U.Chicago, MPI Munich, AEI Golm, NBI Copenhagen, U.Maryland, U.Berkeley, Ohio State University and Northeastern university for hospitality where portions of the work reviewed, were completed.

\end{document}